\documentclass[a4paper,pra,twocolumn,altaffillsymbol,showpacs]{revtex4}
\usepackage{graphics}
\usepackage{graphicx,latexsym,color}
\begin{document}
%\hbadness = 10000
\title{A compact dual atom interferometer gyroscope based on laser-cooled rubidium}
\author{T. M\"{u}ller}\altaffiliation[Present address: ]{Nanyang Technological University, 1 Nanyang Walk,
Block 5 Level 3, Singapore 637616, Singapore}
\author{M. Gilowski}
\author{M. Zaiser}
\author{T. Wendrich}
\author{W. Ertmer}
\author{E. M. Rasel}
\affiliation{Institut f\"{u}r Quantenoptik, Leibniz Universit\"{a}t Hannover, Welfengarten 1,
30167 Hannover, Germany}
\begin{abstract}
We present a compact and transportable inertial sensor
for precision sensing of rotations and accelerations. The sensor
consists of a dual Mach-Zehnder-type atom interferometer operated
with laser-cooled $^{87}$Rb. Raman processes are employed to
coherently manipulate the matter waves. We describe and
characterize the experimental apparatus. A method for passing from
a compact geometry to an extended interferometer with three
independent atom-light interaction zones is proposed and
investigated. The extended geometry will enhance the sensitivity
by more than two orders of magnitude which is necessary to achieve
sensitivities better than $10^{-8}\,$rad/s/$\sqrt{\rm Hz}$.
\end{abstract}
\pacs{03.75.Dg, 37.25.+k, 06.30.Gv}
%\authorrunning{T. M\"{u}ller et.al.}

\maketitle
\par
\section{Introduction}
\label{intro} During the last years, atom interferometry has become an
outstanding technique for precision measurements of fundamental
constants~\cite{Wicht02,Bertoldi06} or inertial forces like
accelerations~\cite{Peters99} or rotations~\cite{Gustavson00}. Thanks to
the progress in quantum engineering of cold atoms, the ultimate potential
of matter wave interferometers is still an open
question~\cite{Kasevich02}, and new atom interferometers aiming for
unprecedented sensitivity to be used for improved tests of the fundamental
laws of physics on ground~\cite{Kasevich07} or in microgravity
environment~\cite{Jentsch04,Vogel06,Nyman06} are an exciting focus of
current research. Other activities concern the realization of atomic
quantum sensors employed in metrology or in applied sciences. Fields of
applications are the monitoring of variations of the geopotential or the
earth's rotation~\cite{Schreiber03}. Today, variations of the earth's
rotation are measured globally by Very Long Baseline Interferometry
(VLBI)~\cite{Schlueter07} or locally with large optical ring laser
gyroscopes~\cite{Dunn02}. Optical gyroscopes massively gain in sensitivity
by enlarging the enclosed area up to hundreds of square meters.

In this paper, we present a novel dual atom interferometer for
exploring the potential of cold atoms for miniaturized and stable
high resolution rotation sensors. The sensor is based on a
Mach-Zehnder-type interferometer, where cold atoms are coherently
split, redirected and recombined to interfere by velocity-selective
Raman transitions. Coherent manipulation by Raman processes was
formerly used for atomic gyroscopes using thermal
beams~\cite{Gustavson00} or laser-cooled atoms~\cite{Canuel06}. Our
apparatus aims to combine some of the advantages of the two
aforementioned experiments. On the one hand, the source system
employed for our sensor generates a high effective atomic flux
comparable to thermal beams~\cite{Mueller07}. Moreover, the sensor is
based on spatially separated beam splitters for long baseline
interferometry allowing to strive for a rotational sensitivity of a
few nrad/s/$\sqrt{\rm Hz}$. On the other hand, we employ laser-cooled
atoms started in a molasses, thus permitting a precise control of the
launch parameters~\cite{Canuel06} and allowing for a well-defined
enclosed area in the interferometer. Additionally, it is anticipated
to increase the long-term stability of the rotation measurement by
employing compact gyroscopes with laser-cooled atoms to achieve
resolutions comparable to~\cite{Gustavson00}. This allows for a
highly transportable sensor for comparison campaigns with other
state-of-the-art optical and atomic gyroscopes.

In the description of our novel design of an atomic gyroscope using
ultra-cold rubidium atoms, we briefly sketch the measurement scheme
and then describe the experimental implementation of the
interferometer's key elements and their performance. Finally, we
present elements and techniques for atom-interferometric in situ
diagnostics for such an inertial sensor. In particular, we propose
and investigate a method to pass from a single laser beam
interferometer to a long baseline design.

\section{Mach-Zehnder interferometry based on stimulated Raman transitions for precision inertial sensing}
\label{sec:1}

The principle of the coherent manipulation of atoms by a two-photon
Raman process is described in detail in Ref.~\cite{Berman97}. In the
Raman process, two long-lived hyperfine states $\left|g\right>$ and
$\left|e\right>$ are coupled by two light fields via an intermediate
state $\left|i\right>$ which is off-resonance from the single photon
excitations.

A Mach-Zehnder-type atom interferometer can be implemented with
three such Raman processes separated in time and/or space. The
first one coherently transfers the atoms initially prepared in one
of the two hyperfine states into a coherent superposition of
$\left|g\right>$ and $\left|e\right>$. Furthermore, this so-called
$\pi/2$-pulse generates two spatially separating matter wave
modes, when the two light fields are counterpropagating, similar
to a 50/50 beam splitter in optics. The second Raman process, a
$\pi$-pulse, flips the internal state and the relative momentum of
both matter waves and thus acts like a mirror. Recombination of
the two matter wave modes is achieved by the third Raman process
acting as a coherent beam splitter mixing the two matter wave
modes, similar to the first $\pi/2$-pulse.

In the ideal case, this type of interferometer has two output
ports labelled by the internal states $\left|g\right>$ and
$\left|e\right>$ and different momenta. A varying phase difference
between the matter wave modes leads to a varying population
difference between the two atomic states which is measured by
state-selective fluorescence detection. This means that the
transition probability
$P_{\left|e\right>}=N_{\left|e\right>}/(N_{\left|g\right>}+N_{\left|e\right>})$
of populating the excited state \cite{f1} depends on the relative
phase $\Delta\phi_{\rm{p}}$ of both matter wave modes and the
relative phase $\Delta\phi_{\rm{L}}$ of the light fields between
the first and the second, and between the second and the third
Raman process:
\begin{equation}\label{eq:1}
P_{\left|e\right>} =
\frac{1}{2}(1+C\,\cos({\Delta\phi_{\rm{L}}+\Delta\phi_{\rm{p}}})).
\end{equation}
Here, $C$ is the contrast of the interferometer and
$\Delta\phi_{\rm{L}}=\phi_1-2\phi_2+\phi_3$ is the combined phase
resulting in the interferometer by the interaction of the atom
dipole with the $i$-th Raman beam splitting pulse $(i=1,2,3)$.

Due to its symmetry, the Mach-Zehnder topology is especially
suited for the measurement of inertial forces. In the ideal case,
the phase difference $\Delta\phi_{\rm{p}}$ depends only on the
relative motion of the atoms with respect to the light fields
driving the Raman process. This relative motion is modified by
inertial forces, i.e. rotations and accelerations such that
$\Delta\phi_{\rm{p}}=\Delta\phi_{\rm{rot}}+\Delta\phi_{\rm{acc}}$.

The phase shift induced by rotations
\begin{equation}\label{eq:2}
\Delta\phi_{\rm{rot}}=\frac{2 m}{\hbar}\vec{\Omega}\cdot\vec{A}
\end{equation}
depends on the orientation and size of the area $\vec{A}$ enclosed
by the interferometer, the rotation vector $\vec{\Omega}$, the
mass $m$ of the particle, and Planck's constant $\hbar$. In an
atom interferometer, this phase shift is a factor
$mc^2/\hbar\omega\sim 10^{10}$ bigger than in an optical
interferometer, assuming the wavelength of visible light and the
rubidium mass, as well as an identical enclosed area. This reveals
the high potential of matter wave interferometers as an
alternative to gyroscopes solely based on light waves, despite the
significantly smaller signal-to-noise-ratio (SNR) and enclosed
area.

An acceleration $\vec{a}$ of the freely propagating atoms relative
to the beam splitting light pulses of the interferometer induces a
phase shift given by
\begin{equation}\label{eq:3}
\Delta\phi_{\rm{acc}}=\vec{k}_{\rm{eff}}\vec{a}T^2,
\end{equation}
where $\vec{k}_{\rm{eff}}$ is the effective wave vector of the
Raman laser beams and $T$ the time interval between the different
beam splitting pulses.

\begin{figure*}[tb]
\resizebox{\textwidth}{!} {\includegraphics{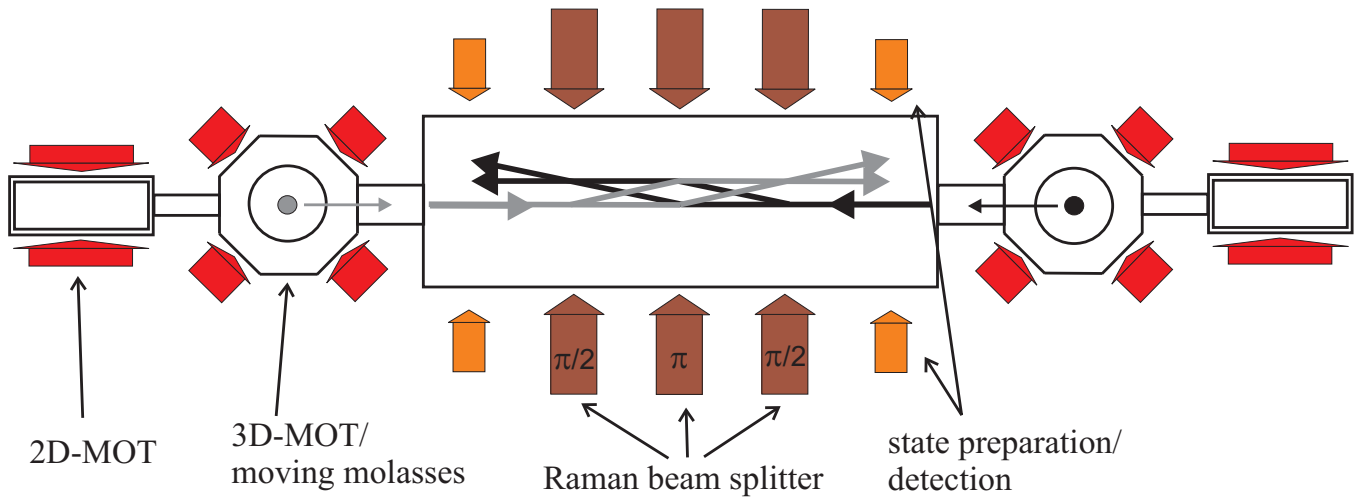}}
\caption{(color online). Schematic of the dual atom interferometer
in top view. The central part shows the interferometry chamber
with three spatially separated atom-light interaction zones. Dual
stage atom sources for the preparation of the cold atomic
ensembles are mounted on both sides of this chamber.}
\label{fig:1}
\end{figure*}

Employing a Mach-Zehnder interferometer as an inertial sensor
requires a distinction between rotations and accelerations. In
order to achieve %a discrimination between rotations and accelerations,
this, we operate the interferometer simultaneously with two atomic
sources %emitting atoms with opposing velocity direction in the
injecting atoms with opposing velocity direction from opposite
sides into the interferometer, as displayed in Fig.~\ref{fig:1}.
This results in a dual interferometer with two equal, but opposite
enclosed areas $\vec{A}_1=-\vec{A}_2$~\cite{Gustavson00}. The sign
of the phase shift induced by rotations therefore differs between
the two interferometers, whereas the phase shift %induced by
due to accelerations is of equal sign, see Eq.~(\ref{eq:2})
and~(\ref{eq:3}). Consequently, subtraction or addition of the two
interferometer signals allows for a discrimination between the two
inertial forces. Additionally, common-mode phase noise of the two
interferometers, for example induced by the Raman laser beams
$\partial\phi_i$, is suppressed by 40 dB due to the utilized
differential measurement scheme~\cite{Canuel07}. The concept of
dual atom interferometry has already been successfully implemented
in two other atomic gyroscopes~\cite{Gustavson00,Canuel06}.

\section{Experimental implementation of atom-optical elements}
\label{sec:2} In this section, we describe the experimental
apparatus of our atomic Sagnac interferometer comprising the
atomic sources, the laser system for the incoherent and coherent
manipulation of the atoms, as well as the state preparation and
detection.

\subsection{The main apparatus}
\label{sec:2a} The differential measurement scheme of the atomic
Sagnac interferometer is reflected by the symmetric design of the
main experimental apparatus shown in Fig.~\ref{fig:1}. It consists of
two identical atomic sources~\cite{Mueller07} generating
counterpropagating pulsed or continuous beams of cold atoms %launched
injected from two opposite sides into the central part, where the
state preparation, the coherent manipulation, and the detection of
the atoms is performed. The vacuum system for the generation and
manipulation of cold atoms is based on custom-designed aluminum
chambers evacuated by a $20\,$l/s ion pump and a titanium
sublimation pump. The main apparatus with an overall length of
only $90\,$cm is placed on a non-magnetic optical breadboard with
a size of $120\times 90\,$cm$^2$. The laser beams for manipulating
the atoms are brought to the experiment via polarization
maintaining (PM) optical fibers, thus decoupling the laser system
from the main apparatus and improving the steady reproducibility
of the atom-light interaction. The main experimental apparatus is
surrounded by a permalloy magnetic shield to suppress stray
fields, and will be put on an active vibration isolation in a
later stage to reduce the accelerational noise from the laboratory
environment.

\subsection{The atomic sources}
\label{sec:2b} The two identical atomic sources will be described
only briefly here, as an extensive description and
characterization can be found in Ref.~\cite{Mueller07}. The atomic
sources consist of a two-dimensional magneto-optical trap (2D-MOT)
loading a subsequent 3D-MOT for pulsed operation of the
interferometer. With the 2D-MOT, we obtain a high loading flux of
more than $5\times10^{9}\,$atoms/s into the 3D-MOT. This allows
for short loading times of the 3D-MOT compared to the measurement
time while still providing atom numbers sufficient to reach a
desirable SNR ($10^8\,$atoms in $20\,$ms loading time).

After loading, the 3D-MOT is switched to a moving molasses
configuration~\cite{Salomon91} launching the atoms into the
central interferometer chamber on flat parabolic trajectories.
Forward drift velocity and vertical velocity can be tuned
independently between $2.5-5\,$m/s and $0-1\,$m/s, respectively,
with a relative uncertainty of $<3\times 10^{-4}$ to realize
perfectly symmetric parabolic trajectories and an optimal spatial
overlap of both interferometers. This permits systematic studies
and optimization of the interferometer sequence with respect to
the suppression of frequency-dependent noise.

Using this technique, the launched atoms, typically started with a
forward velocity of $4.4\,$m/s, reach a temperature of $8~\mu$K.
The precise control of the atomic velocity results in an accuracy
of the enclosed interferometer area sufficient for reaching
resolutions of rotation rates on the order of a few nrad/s with
the completed gyroscope~\cite{Mueller07}.

\subsection{Manipulation of atoms with laser light}
\label{sec:2e} Six diode laser systems are used for the
manipulation of the $^{87}$Rb atoms with light. These are set up
together with further optical elements for frequency stabilization
and beam manipulation on two $60\times90\,$cm$^2$ optical
breadboards. Cooling and trapping of the atoms is performed with
two commercial high-power diode laser systems locked close to the
$\left|F=2\right>\rightarrow\left|F'=3\right>$ cooling transition.
The light for repumping atoms from $\left|F=1\right>$ to
$\left|F=2\right>$ via optical pumping from
$\left|F=1\right>\rightarrow\left|F'=2\right>$, for fluorescence
detection by driving the transition
$\left|F=2\right>\rightarrow\left|F'=3\right>$, as well as for the
state preparation addressing the transition
$\left|F=2\right>\rightarrow\left|F'=2\right>$ is realized with
self-made external cavity diode lasers (ECDL)~\cite{Ricci95}.\\

\emph{State preparation and detection}\\

The two atomic states used for interferometry are the magnetically
insensitive $m_F=0$-Zeeman states of the hyperfine ground states
$\left|F=1\right>$ and $\left|F=2\right>$ of $^{87}$Rb. After the
launch, the atoms populating all $m_F$-substates of
$\left|F=2\right>$ are spin-polarized into the
$\left|F=1,m_F=0\right>$ state by the following procedure: We
apply a $0.4\,$ms long $\pi$-polarized light pulse resonant with
the $\left|F=2\right>\rightarrow\left|F'=2\right>$ transition. The
atoms are thus optically pumped into the $\left|F=2,m_F=0\right>$
ground state which decouples from the light due to optical
selection rules. During optical pumping, we simultaneously repump
atoms decayed into the $\left|F=1\right>$ state by addressing the
$\left|F=1\right>\rightarrow\left|F'=2\right>$ transition. The
repumping light is maintained $0.4\,$ms longer than the pumping
light to completely depopulate the lower hyperfine level. The
optical pump is followed by a resonant $\pi$-pulse with the Raman
lasers at a frequency difference of about $6.834\,$GHz to transfer
the atoms from the $\left|F=2,m_F=0\right>$ state into the
$\left|F=1,m_F=0\right>$ state. The atoms remaining in the
$\left|F=2\right>$-manifold after the $\pi$-pulse are removed by
the light pressure from a $\sigma^+$-polarized light pulse tuned
in resonance with the closed transition
$\left|F=2\right>\rightarrow\left|F'=3\right>$. The spin
polarization of this preparation scheme is better than $99.9\%$.
The optical pumping leads to a slight heating of the atomic sample
to about $10~\mu$K. In the upgraded version of our Sagnac
interferometer, we plan to combine the internal state selection
with a velocity filter by employing velocity-selective Raman
processes~\cite{Kasevich91}.

The detection of the relative atom number
$N_{\left|e\right>}/(N_{\left|g\right>}+N_{\left|e\right>})$ is
performed in a sequence similar to the fluorescence detection
schemes used in atomic fountain clocks~\cite{Bize99}, where our
sequence is pulsed in time rather than in space. In the first
step, we determine the number $N_{\left|e\right>}$ of atoms in the
state $\left|e\right>=\left|F=2,m_F=0\right>$ by applying a
$0.8\,$ms long nearly resonant detection pulse at a detuning of
$\delta=5\,$MHz on the closed transition
$\left|F=2\right>\rightarrow\left|F=3\right>$. The light beam with
an intensity of $I_{sat}$ \cite{f2} is $\sigma^+$-polarized and
retro-reflected to avoid loss of atoms. The induced fluorescence
is recorded on a calibrated photo diode collecting about $1.8\%$
of the isotropically scattered light. This detection step is
followed by a $0.4\,$ms long repumping pulse on the transition
$\left|F=1\right>\rightarrow\left|F'=2\right>$, thus bringing all
the atoms into the upper hyperfine ground state. In the last step,
we repeat the first detection light pulse and by using the same
photo diode we determine the absolute number of atoms
$N_{\left|g\right>}+N_{\left|e\right>}$. We currently reach a SNR
of 30 with the described detection scheme which is mainly limited
by the
frequency noise of the detection laser.\\

\emph{Coherent beam splitting with Raman processes}\\

The interferometer's light pulse sequence is performed by two
digitally phase-locked Raman lasers driving the hyperfine
transition at $\omega_{\rm{eff}}=\omega_1-\omega_2\approx
6.834\,$GHz. The two lasers are tuned off-resonance from the
transitions $\left|F=1\right>\rightarrow \left|F'=1\right>$ and
$\left|F=2\right>\rightarrow \left|F'=1\right>$ by the variable
detuning of $\Delta=0.4-3\,$GHz.

The two self-made high-power diode laser systems~\cite{Voigt01}
have a total laser power of $700$ and $500\,$mW, respectively.
This is sufficient to reach pulse lengths of as short as several
$\mu$s for the realization of the interferometer pulses (at a
detuning $\Delta$ of $1\,$GHz), even for the rather large
diameters of $30\,$mm used for the Raman beam splitters. This
pulse length has been inferred to be optimal with respect to the
matching of the Doppler-broadened transition with the
Fourier-limited excitation profile of the Raman transition, as
well as with respect to the influence of frequency dependent noise
sources in our interferometer~\cite{Cheinet05}.

\begin{figure}[tb]
\includegraphics[width=8.8cm]{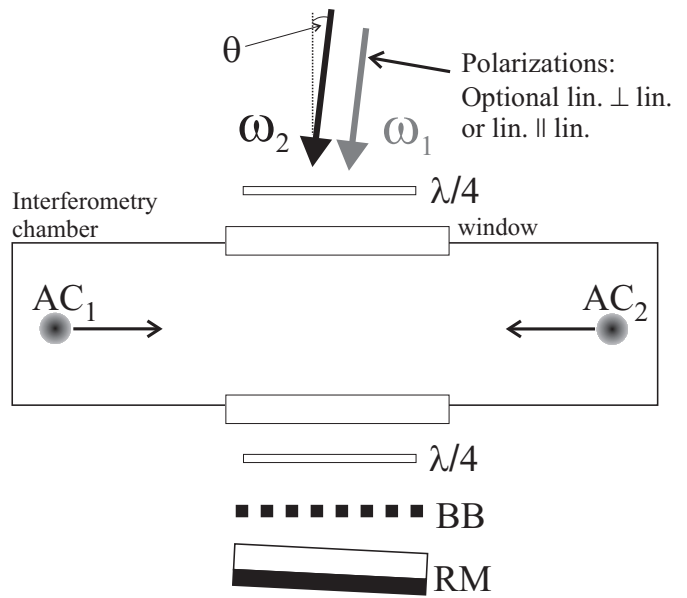}
\caption{Schematic of the optical realization of the Raman beam
splitters. The two atomic clouds (AC$_{1,2}$) are coherently
manipulated by inducing Raman transitions with the two laser beams
of frequency $\omega_1$ and $\omega_2$. The transition can be
induced with equal circular ($\sigma^+-\sigma^+$ or
$\sigma^--\sigma^-$) or crossed linear polarization of the two
lasers. This is used to choose the velocity-selectivity of the
transition by changing the polarization of the incoming Raman
beams. For velocity-unselective transitions, the beams are
absorbed by a beam block (BB) after crossing the atom-light
interaction zone once, while for velocity-selective transitions
they cross the interaction zone twice by using a retro-reflecting
mirror (RM).} \label{fig:1a}
\end{figure}

The phase difference of the two lasers is tightly locked with an
optical phase-locked loop (PLL)~\cite{Santarelli94}. This is
achieved by monitoring the beat note of the two light fields and
comparing the beat frequency and phase difference with a
high-quality crystal oscillator. Our system realizing the PLL is
almost identical to the one described in~\cite{Cheinet05}. We
specify the rms phase error in each of the interferometers caused
by the phase noise of the Raman laser system to be about $1\,$mrad
without common-mode noise suppression. This estimation is based on
the residual phase noise of the Raman laser system and the
weighting function of our upgraded interferometer, where the
weighting function is a measure for the influence of frequency
dependent noise sources on the interferometer's performance. The
interferometer experiences different sensitivities to phase noise
during one experimental cycle, for example the loading time of the
MOT (no sensitivity) compared to the time between the beam
splitting pulses (highest sensitivity).

Both beams are guided to the atom-light interaction zone in the
same PM optical fibre with either identical or crossed linear
polarization for velocity-unselective or velocity-selective Raman
transitions, respectively. A schematic of the optical realization
of the velocity-unselective and velocity-selective Raman beam
splitters, respectively, is depicted in Fig.~\ref{fig:1a}. To
drive velocity-unselective transitions, the two beams are blocked
after passing the interferometer once. To implement
velocity-selective transitions, the two beams pass a
$\lambda/4$-wave plate before and after the atom-light interaction
zone and are then retro-reflected to pass the wave plates and the
interaction zone a second time. In this configuration, only
velocity-selective Raman transitions with opposing laser beam
directions are possible due to optical selection rules. By
aligning the laser beams with a small angle offset of
$\theta\approx 12\,$mrad from perpendicular incidence with respect
to the atomic trajectories, a small Doppler shift can be
introduced in the velocity-selective Raman transition. Thus, one
of the (otherwise degenerate) Raman transitions induced by the two
pairs of counterpropagating laser beam pairs ($\sigma^+-\sigma^+$
or $\sigma^--\sigma^-$) can be selected by %choosing the
compensating the corresponding Doppler detuning by the frequency
difference $\omega_{\rm{eff}}$ of the Raman laser beams. While in
the first case of velocity-unselective Raman transitions only the
photon recoil corresponding to the microwave frequency
$\omega_{\rm{eff}}$ is transferred to the atoms,
velocity-selective Raman transitions will lead to a momentum
transfer of roughly twice the recoil of the optical photon.

For the studies presented in this paper, we used a Raman atom
interferometer with an effective length of about $9\,$mm,
corresponding to a total interferometer time of $2T=2$~ms. The co-
or counterpropagating pair of Raman laser beams is pulsed onto the
atoms, while these cross the Gaussian laser beam profile in the
center of the interferometer chamber. The large beam diameter, as
well as optics with a specified distortion of less than
$\lambda/20$ assure a sufficient quality of the wave front.

\begin{figure}[tb]
\includegraphics[width=8.8cm]{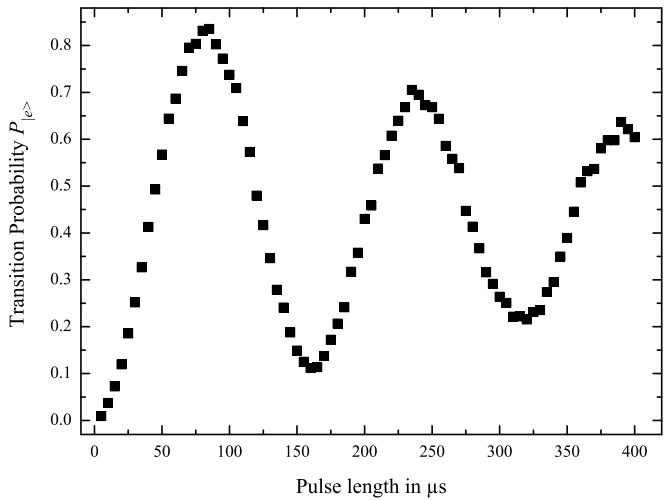}
\caption{Rabi oscillations between the two atomic states used in
the interferometer induced by velocity-unselective Raman
transitions. The oscillations are observed by varying the pulse
length of the Raman transition.} \label{fig:2}
\end{figure}

\section{Atom-interferometric in situ diagnostics for inertial sensing}

\label{sec:3} Characterization and optimization of the atom
interferometer is an iterative process of various in situ
measurements. The flexibility of the Raman process with respect to
the coherent manipulation of the atoms allows for the
implementation of a variety of atom-interferometric tools to
characterize the performance and systematics of the inertial
sensor. This tool box comprises the individual
velocity-unselective or -selective Raman process as the simplest
element to coherently manipulate the atoms, Ramsey-type
experiments to analyze systematics, as well as Mach-Zehnder-type
interferometry to determine the orientation of the setup or
ultimately to measure rotation rates. %In the following sections,

In the following sections, we describe measurements to evaluate
the performance of our sensor including the influence of the
atomic temperature and the effective atomic flux $\Phi_{\rm eff}$.
The effective atomic flux $\Phi_{\rm eff}$ is given by twice the
product of the total number
$N_{\left|g\right>}+N_{\left|e\right>}$ of atoms detected per
shot, the contrast $C$ of the interferometer, and the cycling rate
$\Gamma$ of the sensor:
\begin{equation}\label{eq:PHI}
\Phi_{\rm eff}=2(N_{\left|g\right>}+N_{\left|e\right>})C\Gamma.
\end{equation}
Additionally, we demonstrate tools for proceeding to the finite
interferometer geometry to achieve the ultimate sensitivity.

\subsection{Alignment and contrast of the Mach-Zehnder Interferometer}
\label{subsec:3a}

\begin{figure}
\includegraphics[width=8.8cm]{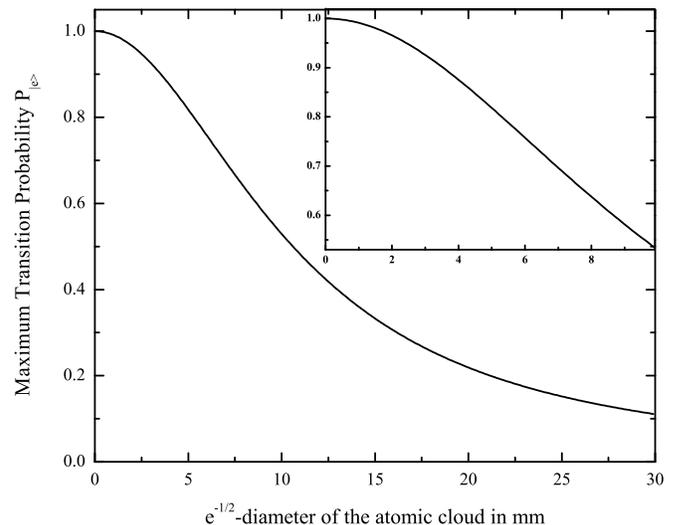}
\caption{Maximum population transfer efficiency as a function of
the diameter of the atomic cloud.} \label{fig:7}
\end{figure}

The achievable contrast $C$ of a Raman atom interferometer is
limited by the parameters of the atomic cloud, such as its spatial
spread in size or its temperature, which have to be compared to
the technical parameters of the matter wave beam splitters. These
are power, detuning, and diameter of the Raman laser beams
determining in combination the length of the $\pi/2$- and
$\pi$-pulses. An upper limit for the achievable contrast can be
inferred from the performance of the individual Raman processes,
i.e. the beam splitting, reflection, and recombination. The
characterization of the velocity-unselective Raman process reveals
the influence of the width of the atomic cloud spread across the
Gaussian intensity profile of the Raman laser beams during the
atom-light interaction. This Gaussian intensity profile leads to
varying Rabi frequencies for the individual atoms manipulated with
the Raman lasers. However, as the length of a beam splitter pulse
is the same for all atoms, this leads to a decrease in the maximum
transition probability being averaged over all atoms. The spatial
spread of the atomic ensemble during the different instants of
atom-light interaction is a function of the distance to the
corresponding atomic source, the drift velocity, the size of the
atomic cloud after preparation, and the temperature of the atomic
ensemble.

\begin{figure}
\includegraphics[width=8.8cm]{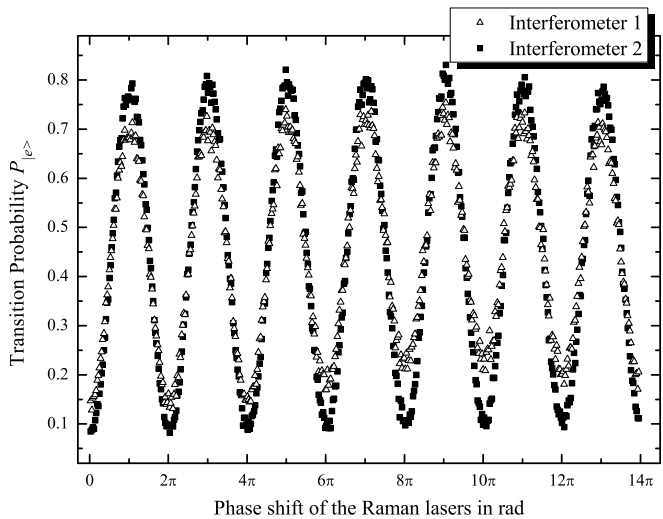}
\caption{Interference fringes in the echo-type sequence
(Mach-Zehnder-type configuration) for velocity-unselective Raman
transitions. To observe the fringes, the phase difference between
the two Raman lasers is scanned between the $\pi$- and the last
$\pi/2$-pulse of the interferometer sequence. The reduced contrast
of interferometer~1 in this particular measurement is caused by a
reduced state preparation efficiency due to technical
imperfections.} \label{fig:5}
\end{figure}

In our setup, the atomic cloud has a $1/\sqrt{e}$-diameter of 4.4~mm
in the middle of the interaction zone which is to be compared to a
$1/e^2$-diameter of the Raman laser beams of 30~mm. This leads to a
maximum population transfer efficiency of $85\%$ with a single
$\pi$-pulse, as shown in Fig.~\ref{fig:2}. The dependance of the
maximum excitation probability, and thus the reflectivity, on the
$1/\sqrt{e}$-diameter of the atomic cloud is displayed in
Fig.~\ref{fig:7}. Further compression of the atomic cloud before the
launch~\cite{Cornell}, a larger diameter of the atom-light
interaction zone and more light power can improve the reflectivity to
about $95\%$. We find that even for the longest flight times of the
atoms the influence of the expansion of the ensemble due to its
temperature is smaller than the initial spatial width.

The synchronization of the Raman pulse with the passing of the
atoms through the center of the Gauss-shaped Raman laser beam is
optimized in situ by minimizing the duration of a $\pi$-pulse by
varying the timing of the pulse. Thus, a precise knowledge of the
start location of the atoms with respect to the Raman laser beams
is not necessary. This method permits to align the
$\pi/2-\pi-\pi/2$-sequence of the dual interferometer with respect
to the atomic parabolas with perfect symmetry.

Fig.~\ref{fig:5} shows a combination of all steps of coherent
manipulation to a $\pi/2-\pi-\pi/2$-sequence, resulting in a
contrast of $72\%$ in our apparatus. The effective atomic flux of
the interferometer is 2.2$\times 10^{7}$~atoms/s. The difference
in contrast between the two atom interferometers is caused by an
imperfect state preparation at the entrance of the
interferometers. The measures for improving the reflection
efficiency of the $\pi$-pulse discussed above would result in a
contrast of the interferometer of more than $90\%$ which may even
be enhanced by an adaption of the length of the beam splitting
pulses, thus compensating the spatially varying Rabi frequency due
to the Gaussian intensity profile of the laser beams over the
whole interferometer region~\cite{Gilowski08}.

\begin{figure}
\includegraphics[width=8.8cm]{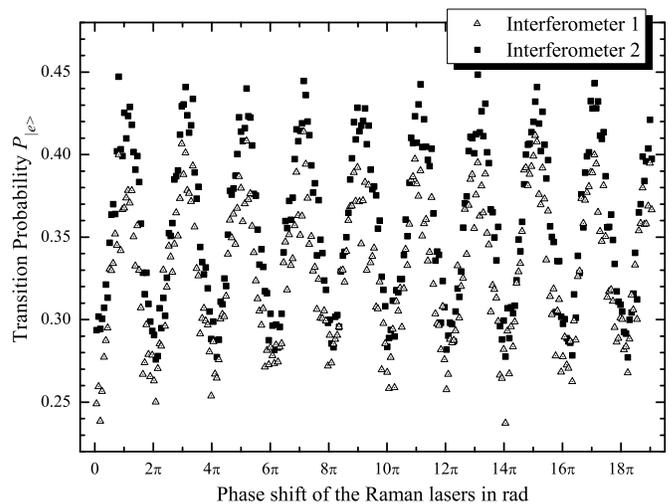}
\caption{Interference fringes in the Mach-Zehnder-type
configuration for velocity-selective Raman transitions. The
oscillations of the transition amplitude are induced as explained
in the caption of Fig.~\ref{fig:5}.} \label{fig:6}
\end{figure}

As shown in Fig.~\ref{fig:6}, the atomic temperature, or more
precisely the velocity dispersion along the Raman laser beams is
much more critical in the case of velocity-selective Raman
transitions. Usually, the frequency width of the resonance of the
Raman process is Fourier-limited by the pulse duration and is by
this means adapted to the atomic temperature. The maximum
transition probability amounts to a fraction of $33\%$ of the
$10\,\mu$K cold atomic ensemble in our apparatus for the
individual velocity-selective Raman process with a pulse duration
of as short as $10\,\mu$s. This leads to a theoretical contrast of
the inertial sensor of $20\%$. The observed contrast shown in
Fig.~\ref{fig:6} is even further reduced due to vibrational noise
which has to be shielded by an active noise cancellation in the
future. The effective atomic flux in the Mach-Zehnder
interferometer is 6$\times 10^{6}$~atoms/s. A contrast close to
the velocity-unselective case can in principle be achieved by a
further reduction of the atomic temperature down to $1.5\,\mu$K or
by adequate velocity filtering~\cite{Kasevich91}.

The presented temporal Mach-Zehnder-type interferometer can be
used to precisely align the long baseline interferometer
perpendicular to local gravity, as well as to optimize the
relative orientation of the spatially separated beam splitters
with respect to each other. For this procedure, each of the
individual Raman sequences has to be replaced by a full temporal
Mach-Zehnder sequence. We demonstrated this method in the central
viewport of our interferometer chamber. Even for a short total
interrogation time of up to $2T=4\,$ms we can obtain a sensitivity
of $2\times 10^{-3}\,$m$/$s$^2/\sqrt{\rm Hz}$ for accelerations
and $2\times 10^{-4}\,$rad$/$s$/\sqrt{\rm Hz}$ for rotations in
our dual atom interferometer.

The intrinsic atomic accelerometer can be used to monitor the
orientation of our atomic Sagnac sensor with respect to local
gravity by combining measurements in the vertical and horizontal
direction with a local tide model. Monitoring the earth's rotation
rate with a resolution of $10^{-9}\,$rad$/$s corresponds to a
control of the angle between the area normal vector of the atom
interferometer and the local direction of gravity to about
$1\,\mu$rad.

\begin{figure}[tb]
\includegraphics[width=8.8cm]{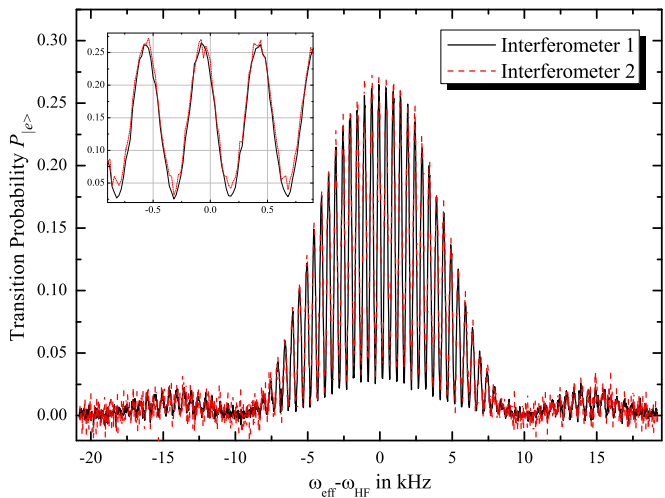}
\caption{(color online). Typical interference pattern for the
Ramsey-type interferometer. The observed oscillations of the
transition amplitude are induced by scanning the frequency
difference $\omega_{\rm{eff}}$ between the two lasers driving the
Raman transition close to the hyperfine splitting
$\omega_{\rm{HF}}=6.8346826\,$GHz.} \label{fig:3}
\end{figure}

\subsection{Control and evaluation of systematic effects by Ramsey-type
measurements}

Besides inertial forces, the Mach-Zehnder interferometer is also
sensitive to changes in the applied frequency difference of the
Raman lasers from the atomic resonance frequency which can e.g. be
perturbed by magnetic or electric fields. Therefore, these effects
have to be evaluated systematically to minimize undesired phase
shifts caused by frequency-changing effects, such as the Zeeman
effect. Contrary to the Ramsey-type measurement employed in atomic
clocks, the sensitivity to frequency changes in the Mach-Zehnder
interferometer does not scale with the time $T$ between the
pulses, but with the length of the pulses $\tau$. We therefore
investigate frequency shifts with Ramsey-type interferometry based
on velocity-unselective Raman processes. This provides valuable
information on the control of systematic shifts and their
contribution to the short-term sensitivity of the inertial sensor
due to fluctuations of the relevant parameters. The signal of the
Ramsey-type interferometer, i.e. the population of the two
hyperfine ground states separated by an energy spacing of $\hbar
\omega_{\rm HF}$ is sensitive to shifts of the frequency
difference $\omega_{\rm eff}$ between the Raman lasers:
\begin{equation}\label{eq:4}
P_{\left|e\right>}\propto (1+\cos
(\omega_{\rm{HF}}-\omega_{\rm{eff}})T).
\end{equation}
Here, $T$ is the separation of the two Raman pulses in time which
is usually significantly larger than the duration of a single
Raman pulse (some ms compared to some $10\,\mu$s). Shifts of the
atomic resonance frequency are inferred by the determination of
the position of the central interferometer fringe while scanning
the frequency difference $\omega_{\rm eff}$ between the Raman
laser pair across the resonance, as shown in Fig.~\ref{fig:3}.

\begin{figure}[tb]
\includegraphics[width=8.8cm]{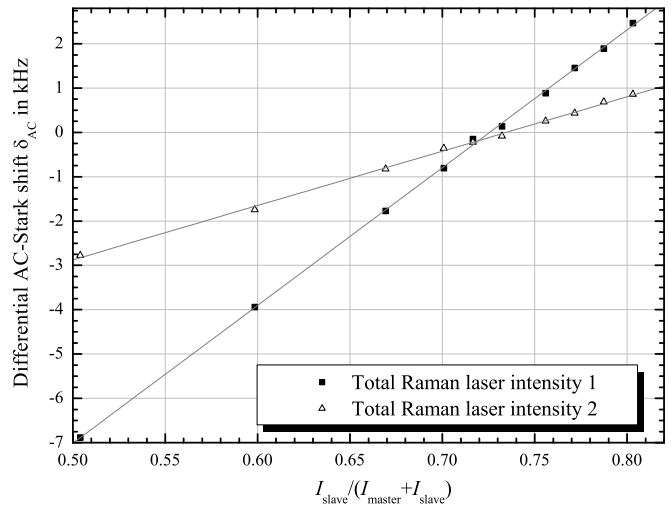}
\caption{Measured differential AC-Stark shift $\delta_{\rm AC}$ of
the atomic resonance as a function of the intensity ratio of the
two Raman lasers for constant total laser intensity. The linear
dependance of $\delta_{\rm AC}$ is shown for two different total
Raman laser intensities.} \label{fig:4}
\end{figure}

We investigated the differential AC-Stark shift, as well as the
Zeeman shift with the Ramsey-type configuration. As the
single-photon AC-Stark shifts of the two atomic levels depend
differently on the intensity of the corresponding Raman laser
($I_{\rm master}$ and $I_{\rm slave}$, respectively), one can find
a ratio of both intensities for which the differential AC-Stark
shift cancels~\cite{Gustavson00}. This ratio is determined by
varying the individual laser intensities for a fixed total
intensity resulting in a linear frequency shift of the resonance,
as shown in Fig.~\ref{fig:4}. With two series of such
measurements, we infer the optimum intensity ratio up to an
uncertainty corresponding to a residual AC-Stark shift
$\delta_{\rm {AC}}$ of $150\,$Hz limited by the power stability of
the Raman lasers during the measurement cycle. This corresponds to
a phase shift of about 13~mrad in our interferometer which should
be suppressed in the differential measurement scheme for slowly
varying intensities of the Raman lasers.

The interferometer is operated with a small magnetic bias field of
typically $400\,$mG to lift the degeneracy of the $m_F$-substates
and to establish a quantization axis for the spin-polarized
ensemble. This necessitates the precise determination of the
second order Zeeman effect and possible fluctuations in the
magnetic bias field. The Ramsey-type measurements show that the
frequency shift due to the quadratic Zeeman effect can be
controlled to about $\pm 1.4\,$Hz. With this in situ measurement,
we determine a reduction of unwanted external magnetic fields in
the interferometer to about 10~mG corresponding to a suppression
of the earth's magnetic field by the permalloy magnetic shield of
at least a factor of 50.

\section{Conclusion and Outlook}
\label{sec:4} In this paper, we have presented a novel design of a
compact dual atom interferometer for high resolution rotation
sensing. The apparatus aims to combine advantages of other
state-of-the-art sensors~\cite{Gustavson00,Canuel06}, such as high
atomic flux equivalent to thermal beams and long baseline
interferometry, but with the advantages of cold atoms. We have
presented different measurement schemes for in situ analysis and
optimization of the dual atom interferometer including the
measurement of systematic effects, such as the Zeeman and the
AC-Stark shift. We indicated a concept to pass from a temporal
sequence with one spatial atom-light interaction zone to long
baseline atom interferometry. The measurement of the sensor's
inclination with respect to local gravity also permits to align
the surface orientation of the sensitive axis of the sensor with
the earth's rotation axis with the required accuracy.

The upgrade of the sensor to a long baseline interferometer will
enhance the sensitivity by more than two orders of magnitude
compared to the temporal Mach-Zehnder geometry realized so far.
The present resolution of phase shifts of $100\,$mrad is mainly
limited by vibrations and technical noise sources of the detection
laser. We expect a reduction of this phase uncertainty down to
$1\,$mrad due to several improvements, like a reduction of the
detection noise by using novel laser
systems~\cite{Baillard06,Gilowski07} with a linewidth two orders
of magnitude lower than the previous system used, a reduction of
the atomic temperature by a factor of 4, as well as an enhancement
of the contrast of the interferometer by a velocity-selective
filter for the atoms participating in the Raman process.

By combining all the described measures, a short-term resolution
of rotation rates of several nrad$/$s for one second of
measurement time comes into reach. This will be an interesting
regime to investigate the potential of differential atom
interferometry and the limitations imposed on the suppression of
common-mode noise or the stability of the setup. These studies
will provide insight on the fundamental limitations of atom
interferometers and their potential for miniaturizing today's
active ring laser gyroscopes searching for relative variations of
the earth's rotation rate of 1 part in $10^7$.

\begin{acknowledgements}
We thank Ch. Jentsch for his contributions during the early stage of
the experiment, G. Santarelli and D. Chambon for their support
regarding the Raman laser system, and F. Pereira Dos Santos for
valuable discussions. This work was supported in part by the
Deutsche Forschungsgemeinschaft (SFB 407) and in part by the
European Union (Contr. No. 012986-2 (NEST), FINAQS). M.G. would like
to thank the Max-Planck-Gesellschaft for financial support.
\end{acknowledgements}
%
% BibTeX users please use
% \bibliographystyle{}
% \bibliography{}

\end{document}